\documentclass{aa}  

\usepackage{graphicx}
\usepackage[dvipsnames]{xcolor}
\usepackage{svg}
\usepackage{txfonts}
\usepackage{calrsfs}
\usepackage{upgreek}
\usepackage{float}
\newcommand{\greenhyperref}[2]{\hypersetup{linkbordercolor=green}\hyperref[#1]{#2}\hypersetup{linkbordercolor=red}}
\DeclareMathAlphabet{\pazocal}{OMS}{zplm}{m}{n}

\newcommand{\citeyearless}[1]{\citeauthor{#1} \citeyear{#1}}
\usepackage{hyperref}
\begin{document}

   \title{Projection-angle effects when ``observing'' a turbulent magnetized collapsing molecular cloud. {II}. Magnetic field}

   \author{A. Tritsis
          \inst{1},
          S. Basu\inst{2,3}
          \and
          C. Federrath\inst{4,5}
          }

   \institute{Institute of Physics, Laboratory of Astrophysics, Ecole Polytechnique F\'ed\'erale de Lausanne (EPFL), \\ Observatoire de Sauverny, 1290, Versoix, Switzerland \\
              \email{aris.tritsis@epfl.ch}
         \and
             Department of Physics and Astronomy, University of Western Ontario, London, ON N6A 3K7, Canada
         \and
             Canadian Institute for Theoretical Astrophysics, University of Toronto, 60 St. George, St., Toronto, ON M5S 3H8, Canada
         \and
             Research School of Astronomy and Astrophysics, Australian National University, Canberra, ACT 2611, Australia
        \and
             Australian Research Council Centre of Excellence in All Sky Astrophysics (ASTRO3D), Canberra, ACT 2611, Australia}
   \date{Received date; accepted date}
   \titlerunning{Probing the projection angle of the magnetic field}
   \authorrunning{Tritsis et al.}
 
  \abstract
   {Interstellar magnetic fields are thought to play a fundamental role in the evolution of star-forming regions. Polarized thermal dust emission serves as a key probe for understanding the structure of the plane-of-the-sky component of the magnetic field in such regions. However, inclination effects can potentially significantly influence the apparent morphology of the magnetic field and lead to erroneous conclusions regarding its dynamical importance.}
   {Our aim is to investigate how projection-angle effects impact dust polarization maps and to explore new ways for accessing the inclination angle of the mean component of the magnetic field with respect to the plane of the sky.}
   {We post-processed a 3D ideal magnetohydrodynamic simulation of a turbulent collapsing molecular cloud at a central density of $10^5~\rm{cm^{-3}}$, when the cloud has flattened perpendicular to the mean magnetic field. We produced synthetic dust polarization measurements under various projection angles, ranging from ``face-on'' (i.e., viewed along the mean magnetic field direction) to ``edge-on'' (perpendicular to the mean magnetic field direction). Additionally, we used synthetic position-position-velocity (PPV) data cubes from the CO ($J = 1\rightarrow0$) transition, presented in a companion paper.}
   {The projected magnetic-field morphology is found to be highly affected by the projection angle with the hourglass morphology being clearly visible only for projection angles close to ``edge-on.' 'We find that the direction of the apparent ``flow'' between successive velocity channels in the simulated PPV data cubes shows an increasing correlation with the synthetic dust polarization observations, as the cloud is observed closer to an ``edge-on'' orientation. Based on this property, we have developed a new method to probe the inclination angle of the magnetic field relative to the plane of the sky. We validated our approach by generating additional synthetic data (PPV cubes and polarization maps) at an earlier stage of the cloud's evolution. We demonstrate an excellent quantitative agreement between the derived inclination angle and the true observational angle. We note that our method is relevant only for collapsing clouds.}
   {}


   \maketitle


\section{Introduction}\label{intro}

The importance of the magnetic field in star formation and the evolution of molecular clouds has been fully recognized from both the theoretical and the observational perspectives (e.g., see \citeyearless{1999ASIC..540..305M}; \citeyearless{2007ARA&A..45..565M}; \citeyearless{2012ApJ...761..156F}; \citeyearless{2014prpl.conf...77P}; \citeyearless{2019FrASS...6....5H}; \citeyearless{2022FrASS...9.9223M}; \citeyearless{2023ASPC..534..317T}; \citeyearless{2023ASPC..534..193P} and references therein). However, the exact significance of the magnetic field relative to gravity and/or turbulence, whether statistically or on a cloud-by-cloud basis, still remains a subject of active debate (e.g., \citeyearless{2012ARA&A..50...29C}; \citeyearless{2014prpl.conf..101L}; \citeyearless{2019FrASS...6....7K}).

Observations of hourglass morphologies in the magnetic field are often regarded as compelling evidence for its dynamical importance, as such morphologies are consistent with collapsing models of magnetized cores (e.g., \citeyearless{1976ApJ...207..141M}; \citeyearless{1993ApJ...415..680F}; \citeyearless{2010MNRAS.408..322K}; \citeyearless{2023MNRAS.521.5087T}). Provided there is at least partial coupling between the magnetic field and the gas, such hourglass morphologies should arise regardless of whether a cloud is initially super- or sub-critical, but the degree of ``pinching'' will be dependent on both the mass-to-flux ratio and the ionization fraction \citep{1966MNRAS.133..265M, 2009NewA...14..221B}. In highly turbulent clouds, this hourglass pattern does not emerge as prominently, as turbulent motions lead the build up of a significant turbulent component of the magnetic field (e.g., see \citeyearless{2017ApJ...838...40M}; \citeyearless{2018MNRAS.480.3916M}; \citeyearless{2018MNRAS.474.5122K}; \citeyearless{2020MNRAS.498.1593B}; \citeyearless{2020MNRAS.499.2076S}; \citeyearless{2021MNRAS.503.5425B}).

Hourglass morphologies in the magnetic field have been observed using submillimeter polarimetric observations in both low-mass (e.g., \citeyearless{2006Sci...313..812G}; \citeyearless{2011A&A...535A..44F}; \citeyearless{2013ApJ...769L..15S}) and high-mass cores \citep{2009Sci...324.1408G, 2014ApJ...794L..18Q, 2024ApJ...972L...6S},  even down to scales of $\le1000$ au \citep{2019A&A...630A..54B}. However, this pattern is not universally seen across observational studies (e.g., see \citeyearless{2024ApJ...963L..31H}; \citeyearless{2019FrASS...6....3H}).

Projection-angle effects can potentially severely complicate the interpretation of infrared and submillimeter polarization observations. Such effects can lead to a distorted view regarding the overall morphology of the magnetic field and its alignment with the major or minor axes of a cloud, as projected on the plane of the sky \citep{2000ApJ...540L.103B, 2012ApJ...761...40K, 2020ApJ...899...28D}. These complications can also lead to erroneous estimates regarding the energetics of the cloud, given that the strength of the magnetic field is estimated based on its projected morphological structure (e.g., \citeyearless{1951Phys.Rev....81...890}; \citeyearless{1953ApJ...118..113C}; \citeyearless{2021A&A...647A.186S}). While statistical arguments regarding the inclination angle (e.g., \citeyearless{2004mim..proc..123C}) can provide insights on average properties across large samples, they cannot be directly applied to individual clouds. Additionally, studies that explicitly probe the inclination angle of the magnetic field relative to the line of sight (LOS) are limited (e.g., \citeyearless{2000ApJ...529..925C}, \citeyearless{2017ApJ...848..110K}, \citeyearless{2019ApJ...874...89C}, \citeyearless{2020ApJ...891...55K}).

Currently, there are three groups of available methods for accessing the inclination angle of the magnetic field in each individual cloud. The first one relies on a combination of the LOS (usually through Zeeman measurements) and plane-of-sky (POS) components of the field (\citeyearless{1991ApJ...373..509M}, \citeyearless{2019ApJ...873...38T}, see also \citeyearless{2022FrASS...9.0027T} and references therein). In the second approach, the inclination of the cloud and its magnetic field is accessed by comparing different models with observations (e.g., \citeyearless{2000ApJ...529..925C}, \citeyearless{2008A&A...490L..39G}, \citeyearless{2022ApJ...936...29B}). Finally, other methods rely on the depolarization due to the projection angle \citep{2019MNRAS.485.3499C, 2023MNRAS.519.3736H}.

In this study, we explore how the inclination angle influences the resulting ``observed'' morphology of the magnetic field and present the method we developed to access this projection angle. Our paper is organized as follows. In Sect.~\ref{numer}, we briefly describe the details of the numerical simulation used, the radiative transfer simulations presented in \greenhyperref{paperI}{Paper I}, and the methodology followed for producing synthetic polarization maps. In Sect.~\ref{mockobs}, we present synthetic dust polarization observations of the cloud under various projection angles. In Sect.~\ref{projangleBfield} we develop a method to probe the projection angle of the magnetic field with respect to the POS based on the correlation between dust emission maps and the apparent ``flow'' in position-position-velocity (PPV) data cubes. We benchmark our method and discuss potential caveats in Sects.~\ref{benchmarking} and ~\ref{cavs}, respectively. Finally, in Sect.~\ref{discuss}, we summarize our most important results.

\section{Numerical methods}\label{numer}

\subsection{Setup and initial conditions}

In \greenhyperref{paperI}{Paper I,} we performed an ideal magnetohydrodynamic (MHD) chemo-dynamical simulation of a turbulent collapsing molecular cloud. In this simulation, we follow the evolution of 115 species, using an ``on-the-fly'' approach. Here, we briefly summarize the main properties of these calculations and refer to the aforementioned study for a detailed description of our numerical methods. 

We simulated an isolated (i.e., open boundary conditions), isothermal (T = 10 K) cloud with a total mass of $\sim$240 M\textsubscript{\(\odot\)} using the astrophysical  \textsc{FLASH} code \citep{2000ApJS..131..273F, 2008ASPC..385..145D}. The cloud was super-critical with a mass-to-flux ratio (in units of the critical value for collapse; \citeyearless{1976ApJ...210..326M}) of 2.3. The magnetic field was initially uniform and oriented along the $z$ axis of our simulation box. We included turbulent initial conditions using the publicly-available code \textsc{TurbGen} \citep{2010A&A...512A..81F, 2022ascl.soft04001F}, with an initial velocity power spectrum $\propto k^{-2}$. The sonic and Alfv\'enic Mach numbers were $\mathcal{M}_s\sim3$ and $\mathcal{M}_A\sim1.2$, respectively.

We post-processed this simulation at a ``central'' number density of $10^5~\rm{cm^{-3}}$ (corresponding to 1.2 times the free-fall time) with the multilevel, non-local thermodynamic equilibrium (non-LTE) radiative transfer code \textsc{PyRaTE} \citep{2018MNRAS.478.2056T, 2024A&A...692A..75T}. We produced synthetic spectral line observations in the form of PPV data cubes of the $J = 1\rightarrow0$ transition for four species; CO, $\rm{HCO^+}$, HCN, and $\rm{N_2H^+}$. For each of the species considered, PPV cubes were produced for the following angles between the LOS and the mean component of the magnetic field: 0$^\circ$, 22.5$^\circ$, 45$^\circ$, 67.5$^\circ$, and 90$^\circ$.

\subsection{Synthetic polarization observations}\label{mockdustmeth}

To create synthetic dust-polarization observations, we follow \citeauthor{2019MNRAS.490.2760K} (\citeyear{2019MNRAS.490.2760K}; see also \citeyearless{2018MNRAS.474.5122K} and \citeyearless{2005ApJ...631..361C}). Here, we summarize the main methodology and refer the interested reader to the aforementioned studies. In column-density units, the Stokes parameters are
\begin{subequations}\label{eq:stokespars}
\begin{equation}\label{stokesI}
I = \int n \Bigg(1- p_0\Bigg(\frac{B_\xi^2 + B_\eta^2}{B^2} - \frac{2}{3} \Bigg)\Bigg) ds,
\end{equation}

\begin{eqnarray}\label{stokesQ}
Q = p_0 \int n \Bigg(\frac{B_\eta^2 - B_\xi^2}{B^2}\Bigg) ds,
\end{eqnarray}

\begin{eqnarray}\label{stokesU}
U = p_0 \int n \Bigg(\frac{2 B_\xi B_\eta}{B^2}\Bigg) ds,
\end{eqnarray}
\end{subequations}
where $n$ is the number density, $B_\xi$ and $B_\eta$ are the components of the magnetic field on the celestial sphere, and $s$ is the length along the LOS. Both $n$ and the magnetic field components $B_\xi$ and $B_\eta$ are functions of position along the LOS, and their variation is accounted for in the integration along the LOS. In Eqs.~\ref{eq:stokespars}, $p_0$ is the polarization efficiency. To create our synthetic dust-polarization observations, we use a varying polarization efficiency given by
\begin{equation}\label{poleff}
p_0 = 0.2577\Big(\frac{\sqrt{a_{max}} - \sqrt{a_{alg}}}{\sqrt{a_{max}} - \sqrt{a_{min}}} \Big)
,\end{equation}
where $a_{max}$ and $a_{min}$ are, respectively, the maximum and minimum dust radii of a grain distribution and $a_{alg}$ is given by
\begin{equation}\label{aalg}
a_{alg} = \frac{(log_{10}n)^3(A_v + 5)}{2800}\mu \rm{m}.
\end{equation}
For $a_{min}$ and $a_{max}$, we followed \cite{2019MNRAS.490.2760K} and adopted 0.005 and 1 $\mu$m, respectively. For the value of the visual extinction that appears into Eq.~\ref{aalg}, we use the pre-computed values of $A_v^{3D}$, as already calculated in the chemo-dynamical simulation (see Eq. 1 from \greenhyperref{paperI}{Paper I}). We note that as our polarization maps were not produced using a full radiative-transfer treatment (e.g., \citeyearless{2016A&A...593A..87R}), they may not achieve the same level of accuracy as the synthetic spectral observations presented in \greenhyperref{paperI}{Paper I}.



\section{Results}\label{Results}

\begin{figure}
\includegraphics[width=1.0\columnwidth, clip]{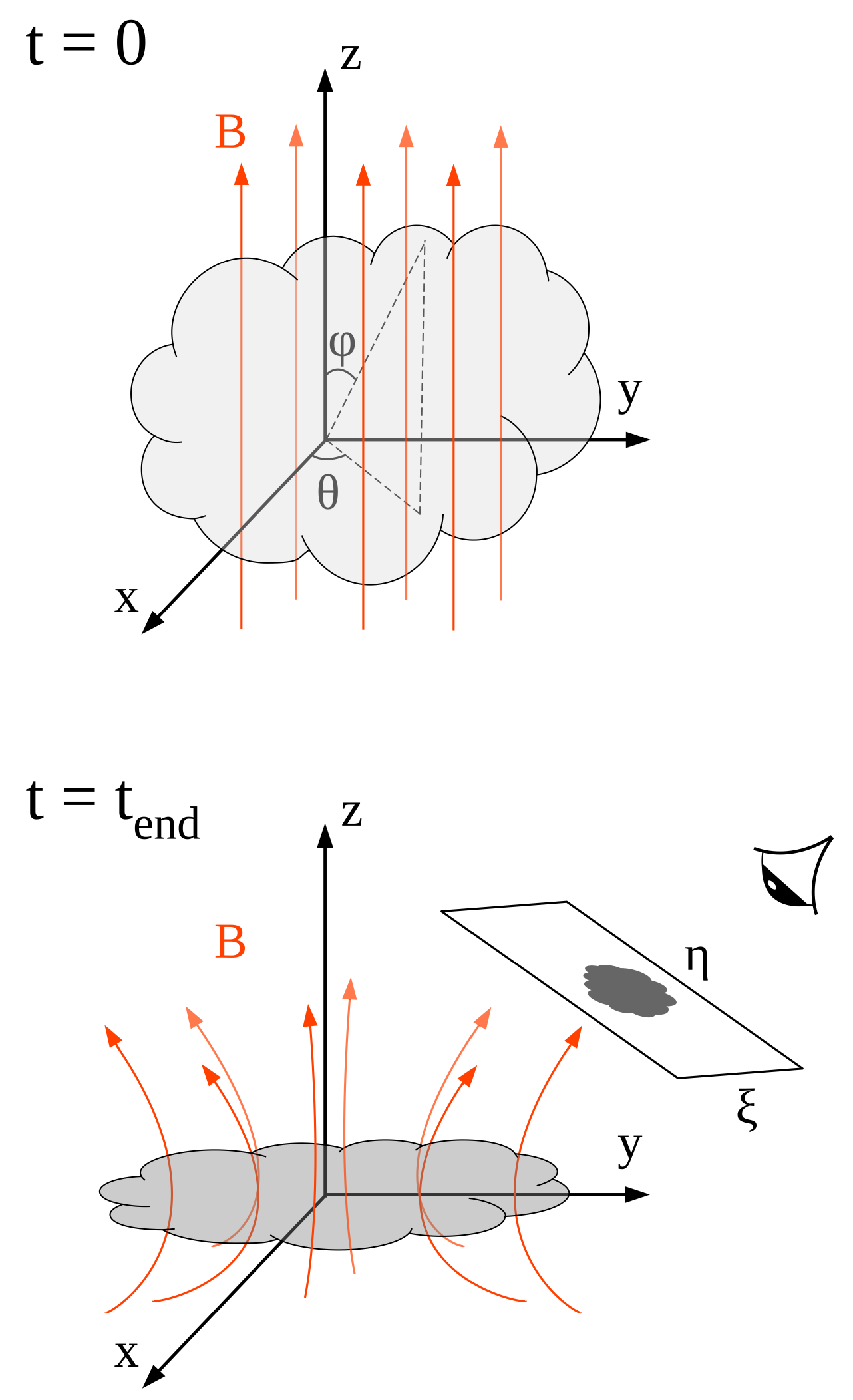}
\caption{Schematic representations of the initial (upper panel) and final (lower panel) configurations of the cloud along with our definitions for the angles $\theta$ and $\phi$, and the directions $\xi$ and $\eta$ on the POS. Orange lines are a schematic representation of the large-scale (i.e., excluding turbulent features) magnetic field. At the time of post-processing the simulation, the cloud has collapsed along the magnetic field, as shown in the bottom panel.
\label{schematic}}
\end{figure}

Throughout our study, we used $\xi$ \& $\eta$ to denote the horizontal and vertical dimensions and directions of the cloud, as projected on the POS. In all figures, we annotated the azimuthal and polar angles ($\theta$ and $\phi$, respectively), which specify the LOS direction, ensuring clarity regarding the projection angle under which the cloud is ``observed.'' Figure~\ref{schematic} provides a schematic representation of the definition of the angles $\theta$ and $\phi$, and the directions $\xi$ and $\eta$. We further illustrate the final configuration of the cloud relative to the large-scale magnetic field (orange lines in Fig.~\ref{schematic}), which is that of a flattened structure, with its short axis oriented parallel to the magnetic field. As is evident from Fig.~\ref{schematic}, when $\phi = \theta = 0^\circ$ the cloud is ``face-on'', and viewed along the $z$ axis. When $\phi = 0^\circ$ and $\theta = 0^\circ$ or $90^\circ$ the cloud is ``edge-on,'' seen along the $x$ or $y$ axes, respectively. For intermediate polar angles (with $\theta = 0^\circ$), the LOS rotates within the $xz$ plane, progressively tilting the view from ``face-on'' to ``edge-on.''

\subsection{Column density and polarization maps}\label{mockobs}

\begin{figure*}
\includegraphics[width=2.075\columnwidth, clip]{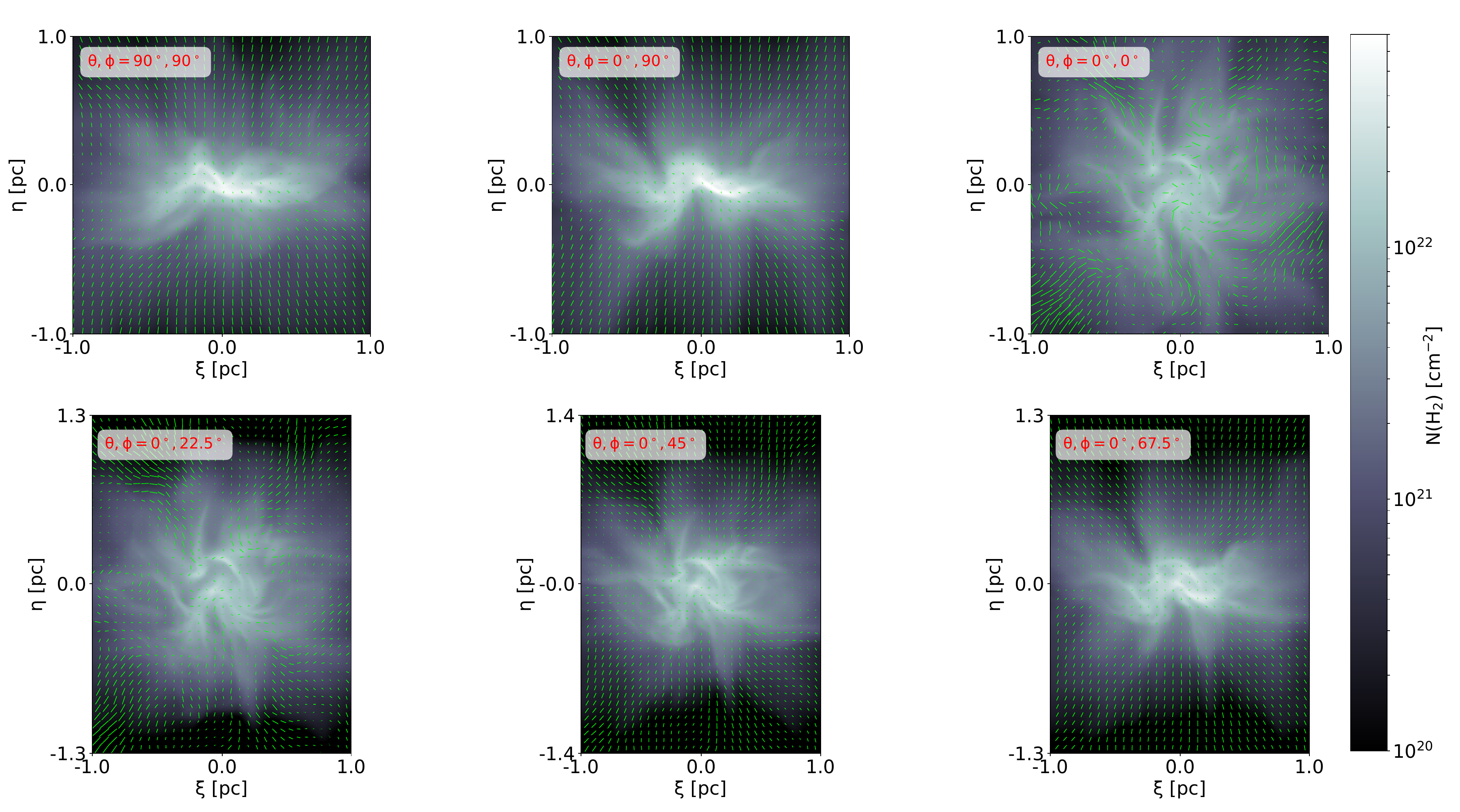}
\caption{Column density maps of our simulated cloud under different projection angles. In the top row, we show the column density along the principal axes of our simulation and in the bottom row, we show the off-axis projections. Green pseudo-vectors are synthetic polarization measurements (see also Fig.~\ref{pcdrel}).
\label{CDs}}
\end{figure*}

In Fig.~\ref{CDs}, we present column-density maps of our simulated cloud under various projection angles. Green pseudo-vectors show synthetic dust-polarization measurements following the approach described in Sect.~\ref{mockdustmeth}. Even though the overall morphology of the cloud is that of a flattened structure perpendicular to the mean component of the magnetic field (see Fig. 1 in \greenhyperref{paperI}{Paper I}), this is not particularly evident in the column density maps under any projection angle. Instead the various filamentary higher-density structures that are formed within the flattened structure ``dominate'' the column density maps in terms of the morphological features. Moreover, as expected, when the projection angle is such that the mean component of the magnetic field primarily lies on the POS (e.g., $\rm{\phi} = 90^\circ$), the column density can be a factor of $\sim$5 higher than when the magnetic field is primarily along the LOS. Interestingly, these are also the projection angles where the cloud appears more filamentary (see upper left and middle panels and lower right panel in Fig.~\ref{CDs}).

The observed magnetic-field morphology is also highly dependent upon the projection angle. Specifically, the hourglass morphology is particularly prominent only for the projection angles where the mean magnetic field lies in (or is close to) the POS. For all other projection angles (especially when $\phi < 67.5^\circ$), the magnetic-field morphology appears significantly more complicated and tangled than what it is in the simulation (see Fig. 1 in \greenhyperref{paperI}{Paper I}). Consequently, such projection effects can significantly influence and bias the results of observational studies aiming to decipher the dynamical importance of the magnetic field, based on its morphological features.

\begin{figure}
\includegraphics[width=1.\columnwidth, clip]{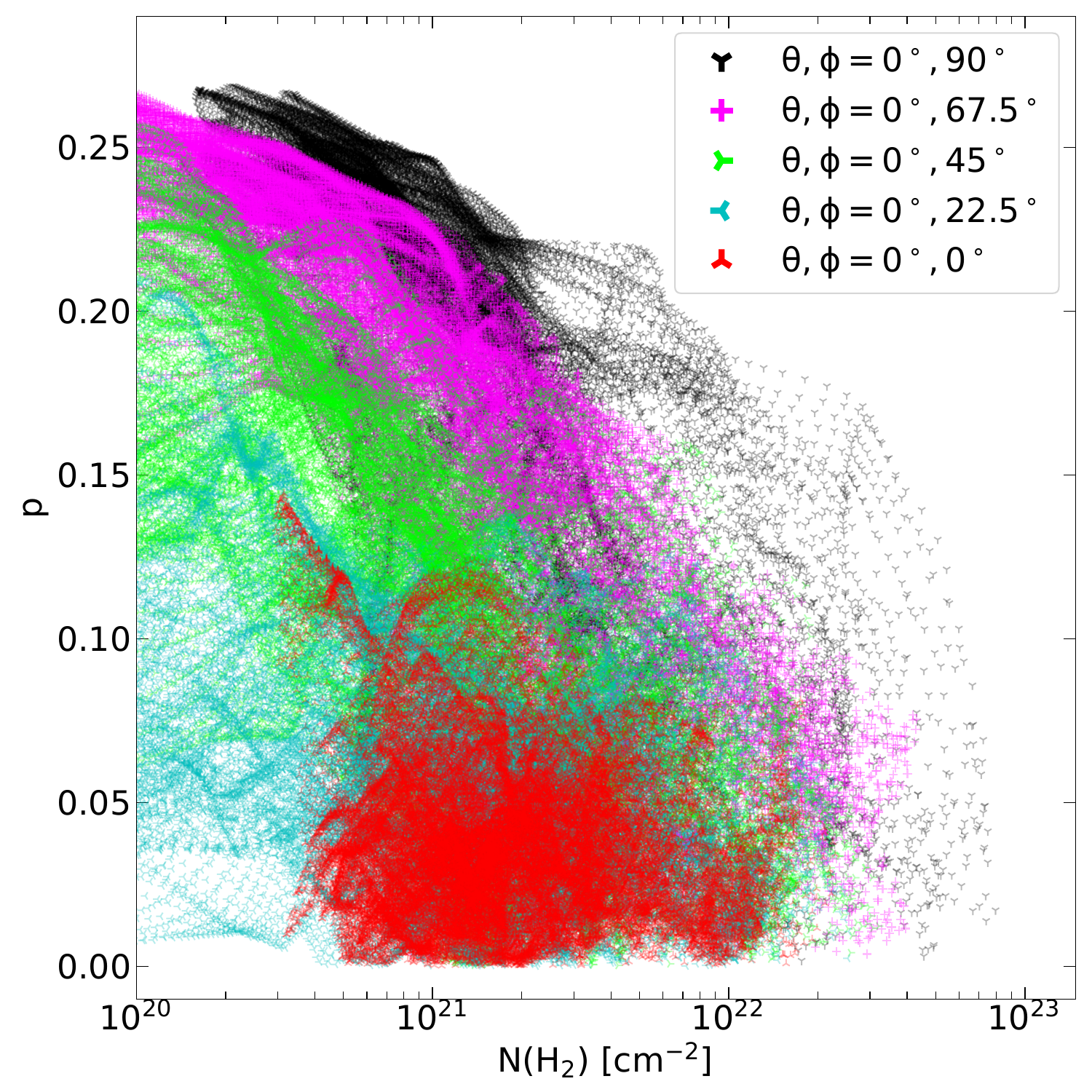}
\caption{Polarization fraction as a function of the $\rm{H_2}$ column density of the cloud for various projection angles. As expected, the polarization fraction is systematically lower, as the angle between the LOS and the mean component of the magnetic field becomes smaller. However, even when observing the cloud ``face-on'' ($\phi$ = $0^\circ$) the polarization is non-negligible, albeit the polarization fraction is probably overestimated for all projection angles.
\label{pcdrel}}
\end{figure}

Fig.~\ref{pcdrel} shows the polarization fraction, defined as
\begin{equation}\label{polfrac}
p = \frac{\sqrt{Q^2+U^2}}{I}
,\end{equation}
as a function of column density, for all projection angles considered here. With the black, magenta, green, cyan, and red points we show our results for polar angles of 90, 67.5, 45, 22.5, and 0$^\circ$, respectively. The polarization fraction systematically decreases when $\rm{\phi}$ decreases. This is to be expected as the mean component of the magnetic field in our simulation is in the $z$ direction. In terms of the values of the polarization fraction, these are overpredicted by $\sim$20\% in comparison to real observations of molecular clouds (e.g., \citeyearless{2015A&A...576A.105P}). This discrepancy can potentially be attributed to the assumptions behind the grain population and polarization cross sections adopted,  which the factor of 0.2577 in Eq.~\ref{poleff} originates from, as well as in the empirical estimate for $\alpha_{alg}$ (see Eq.~\ref{aalg}). As is evident from Fig.~\ref{pcdrel}, the polarization fraction also decreases with column density. This is to be expected given the methodology used to produce the synthetic polarization maps, as in the higher column-density regions, both the visual extinction and the density are higher, leading to a decrease in the polarization efficiency.

\subsection{Correlation between PPV cubes and dust polarization maps}\label{projangleBfield}

One striking feature seen in PPV cubes of real molecular clouds as well as in the synthetic PPV cubes presented in \greenhyperref{paperI}{Paper I} is the presence of an apparent ``flow'' between successive velocity channels (e.g., \citeyearless{2008ApJ...680..428G}). This apparent ``flow'' does not directly correspond to the actual velocity field of the cloud, but is a combination of the LOS velocity between neighboring regions, the molecular abundance distribution, and it is also subject to all other quantities affecting the excitation of the molecule in question. Regardless, in many scenarios this apparent ``flow'' appears to be relatively smooth and examining these motions could provide significant insights regarding the intermittency of turbulence \citep{1995A&A...293..840F, 2008A&A...481..367H}. In this section, we describe how we have taken advantage of this property. We propose a novel approach for analyzing PPV data cubes, which has the potential to unveil the projection angle of the magnetic field. 

\subsubsection{Quantifying the apparent ``flow'' in PPV cubes}

We made use of the Lucas-Kanade optical flow estimator \citep{2005IEEE.1.I137} to trace the direction of the ``flow'' between successive velocity slices in our synthetic PPV cubes. The algorithm functions by assuming that the motion within a local neighborhood of pixels is small and uniform and then solving a system of linear equations of the form $J^{(i, j)}_\xi u_\xi+ J^{(i, j)}_\eta u_\eta + J^{(i, j)}_t = 0$. In the latter equation, ($u_\xi, u_\eta$) are the velocity components in the $\xi$ and $\eta$ directions of the image, respectively, and $J^{(i, j)}_\xi$, $J^{(i, j)}_\eta$, and $J^{(i, j)}_t$ represent the partial derivatives of the image intensity with respect to each spatial coordinate and time. These derivatives are computed for each pixel $i, j$ within the local neighborhood considered. The resulting output from the Lucas-Kanade flow estimator is a vector field ($u_\xi, u_\eta$) that characterizes the displacement of features between successive frames or, in this instance, between successive velocity channels (since the time derivative in our scenario is, in actuality, a derivative over the frequency or velocity channels of the spectral line).

\begin{figure*}
\includegraphics[width=2.\columnwidth, clip]{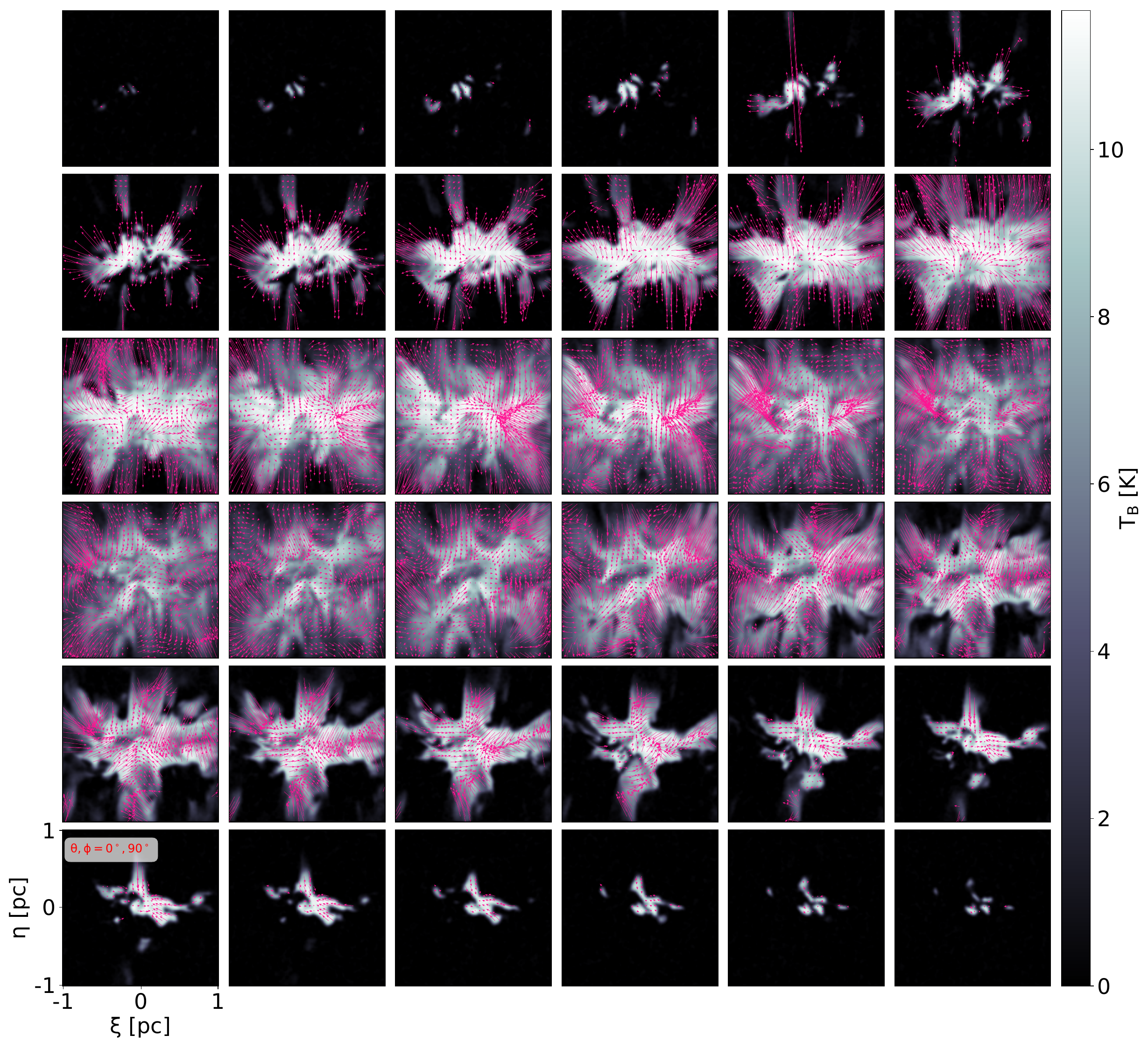}
\caption{Velocity slices from a synthetic $\rm{CO} ~(J = 1\rightarrow0)$ PPV cube in the velocity range $\sim -0.9~\rm{km~s^{-1}}$ (upper left) to $\sim 0.9~\rm{km~s^{-1}}$ (lower right) when the cloud is observed ``edge-on.'' The fuchsia vectors overplotted on top of slice ``n'' mark the direction of the ``flow'' between slice ``n'' and ``n+1.''
\label{FigVecs}}
\end{figure*}

\begin{figure*}
\includegraphics[width=2.\columnwidth, clip]{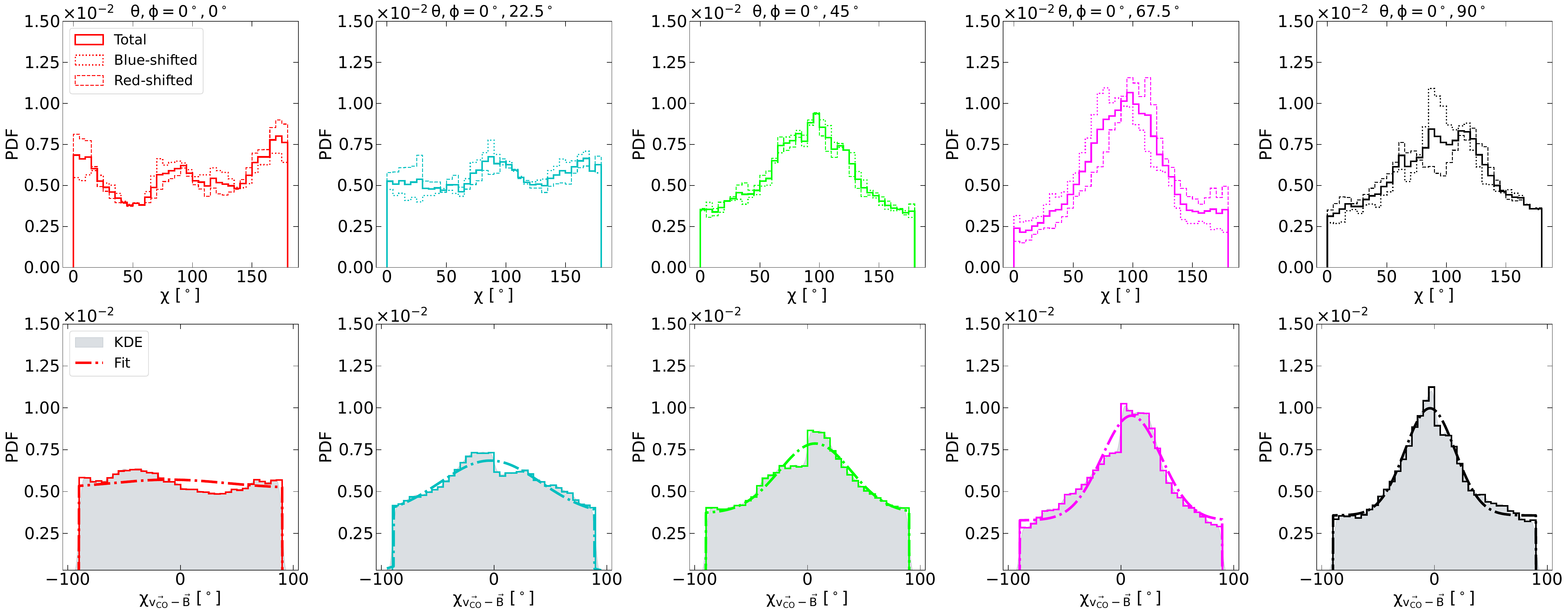}
\caption{Upper row: PDFs of the orientation of the vectors shown in Fig.~\ref{FigVecs} with respect to the horizontal axis of the image (i.e., the $\xi$ direction). For every projection angle, we show the total distribution of angles between all successive velocity slices (solid thick lines), the distribution of angles for the blue-shifted components (dotted lines), and the distribution of angles for the red-shifted components (dashed lines). Lower panels: PDFs of the relative orientation between the vectors shown in Fig.~\ref{FigVecs} and the polarization vectors shown in Fig~\ref{CDs}. The shaded region shows a KDE estimate for each distribution and the dashed-dotted line shows a simultaneous fit of a Gaussian and a uniform distribution to the data.
\label{AngleDists}}
\end{figure*}

In Fig.~\ref{FigVecs}, we show different velocity channels from our synthetic $\rm{CO}$ ($J=1\rightarrow$0) PPV cube presented in \greenhyperref{paperI}{Paper I}, when the cloud is observed ``edge-on'' (i.e., $\phi = 90^\circ$). The PPV cubes for the same transition were produced for all other polar angles examined in the present study. Only the velocity channels in the velocity range [-0.9, 0.9] $\rm{km~s^{-1}}$ are shown, since beyond this range, there is no significant emission (for this particular projection angle). The fuchsia vectors annotated on top of each velocity slice ``n'' mark the direction of the apparent ``flow'' between velocity slices ``n'' and ``n+1''. In order to avoid measuring such vectors in velocity slices and in regions of the cloud where there is no significant emission and, therefore, the direction of the vectors will be dependent exclusively upon random noise, we require that the emission in both velocity slices, ``n'' and ``n+1'', be at least 2.5 times the root mean square (rms) noise in each synthetic cube. We note that, given that the intensity of the lines changes (depending on the inclination angle) and noise is added in such a way that we achieve the same signal-to-noise ratio (S/N) in all synthetic PPV cubes, the rms noise level is different in each PPV cube (see Sect. 2.4 in \greenhyperref{paperI}{Paper I}). To further mitigate the effect of noise, we have also decreased the spatial resolution of the PPV cubes by applying a Gaussian kernel with a standard deviation of one pixel. As made evident upon visually inspecting Fig.~\ref{FigVecs}, the algorithm performs very well in terms of capturing the direction of the ``flow'' and could therefore present a more powerful way of analyzing PPV cubes than just considering the moment maps.

\subsubsection{Connecting the velocity ``flow'' and dust polarization directions}

We  proceed to examine the orientation of these vectors relative to an arbitrary axis, which we chose  as the horizontal axis of the image. This angle is henceforth defined as $\chi$ and the angle of these vectors with respect to the polarization pseudo-vectors (shown in Fig.~\ref{CDs}) is defined as $\chi_{\vec{v_{CO}}-\vec{B}}$. In the upper row of Fig.~\ref{AngleDists} we show the probability density functions (PDFs) of $\chi$ (solid lines) for each inclination angle considered. Dotted and dashed lines show the PDFs of $\chi$ only between blue-shifted and red-shifted velocity components, respectively. In the bottom row of Fig.~\ref{AngleDists}, we show the PDFs of angles between the vectors in Fig.~\ref{FigVecs} and the polarization pseudo-vectors in Fig.~\ref{CDs}. The distributions of angles are shifted from the interval [0$^\circ$, 180$^\circ$] to [-90$^\circ$, 90$^\circ$]. The shaded regions in the bottom row show a kernel-density estimate (KDE) of each distribution and the dash-dotted lines are a simultaneous fit of a uniform and a Gaussian distribution to the data. The fitting is performed to the results from the KDE such that our results are independent of the bin width.

Some trends in Fig.~\ref{AngleDists} are readily visible. Firstly, when the cloud is observed ``face-on'' the distribution of angles with respect to the horizontal axis of the image (upper left panel) resembles a uniform distribution with some weak features close to 0$^\circ$, 90$^\circ$, and 180$^\circ$. The fact that this distribution is close to a uniform distribution can be understood by imagining a perfectly symmetric disk-like collapsing cloud with no turbulent motions and an hourglass magnetic-field morphology as follows. The first features that will be seen in a PPV cube when such a structure is observed ``face-on'' will be at the center of the disk where the LOS curvature of the magnetic field will be zero and the LOS velocity will be larger. In subsequent velocity channels, features will appear progressively farther away from the center, in a circularly symmetric fashion, as these will be the regions with the next highest LOS velocities. An opposite trend, is expected for the red-shifted components. That is an inward ``motion'' between successive channels from the outskirts to the center of the cloud will be observed. By applying the Lucas-Kanade optical flow estimator in such a scenario, the vectors should point radially outwards and inwards in the blue-shifted and red-shifted frequencies, respectively, and their angle distribution with respect to any axis should be close to uniform. 

In the opposite extreme, where this hypothetical cloud is observed ``edge-on'', there should be a preferred direction these vectors should point for a similar reason as in the ``face-on'' case. The regions close to the midplane of the disk, collapsing radially, should appear first in the respective PPV cube followed by regions above and below the midplane, again due to the hourglass morphology of the magnetic field. The vectors tracing the direction of the flow will thus be mostly oriented in the axial direction (i.e., the $\eta$ direction given the definitions followed here) and their orientation, with respect to some arbitrary axis, will exhibit a peak at some angle. 

Even though our simulated cloud does not have a perfect disk-like morphology and we start with initially turbulent initial conditions this trend still holds. Additionally, as evident from Fig.~\ref{AngleDists}, there is a relatively gradual transition from ``face-on'' to ``edge-on'' with intermediate projection angles exhibiting a behavior in between the two extreme cases.

We note here that the behavior seen in $\rm{CO}$ PPV data cubes, is not universal for all molecules modeled in \greenhyperref{paperI}{Paper I}. For $\rm{HCN}$ and $\rm{HCO^+}$, we get a similar qualitative behavior as for $\rm{CO}$, albeit the trend with the inclination angle is somewhat weaker and the distributions are more irregular (not shown here). For $\rm{N_2H^+}$ the trend changes completely, and when the cloud is observed ``edge-on'' the angle distribution between the vectors tracing the ``flow'' and the horizontal axis of the image peaks at 0$^\circ$ and 180$^\circ$. Similarly, with respect to the polarization pseudo-vectors and for an inclination angle of $\phi = 90^\circ$, the PDF of $\chi_{\vec{v_{CO}}-\vec{B}}$ peaks at -90$^\circ$/90$^\circ$. This is to be expected as $\rm{N_2H^+}$ emission for this projection angle is confined in a small dense, elongated region (Fig. 6 in \greenhyperref{paperI}{Paper I} and the upper middle panel in Fig.~\ref{CDs}) and variations in between successive velocity channels are primarily aligned parallel to the horizontal axis. Therefore, for the trend observed in Fig.~\ref{AngleDists} to hold, a molecule probing the bulk of the cloud should be used.

\begin{figure}
\includegraphics[width=1.\columnwidth, clip]{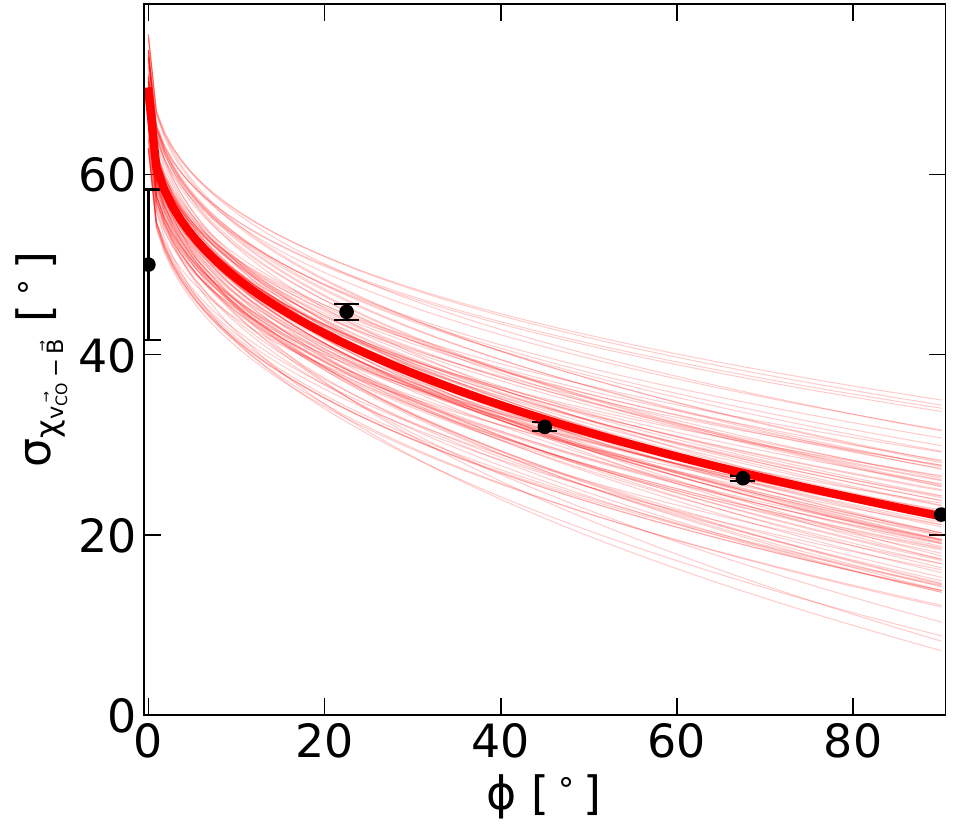}
\caption{Standard deviation of the Gaussian fitted in the bottom row in Fig.~\ref{AngleDists} as a function of the projection angle under which the cloud is observed. The thick red solid line shows the best-fit power-law model (see Eq.~\ref{fittosigma}) obtained using the maximum likelihood estimation and the thin red lines show 100 random samples from the posterior distributions of the parameters.
\label{SigmafPrj}}
\end{figure}

From the analytical fits of the Gaussian and the uniform distribution to the PDFs shown in the bottom row of Fig.~\ref{AngleDists} we now proceed to examine how the standard deviation of the Gaussian distribution correlates with the projection angle. We plot our results in Fig.~\ref{SigmafPrj}. The black points show the standard deviation and the errorbars are the errors from the fit. The data follow the relation:  
\begin{equation}\label{fittosigma}
 \sigma_{\chi_{\vec{v_{CO}} - \vec{B}}} = c_1\phi^\kappa + c_2. 
\end{equation}
We used a Markov chain Monte Carlo (MCMC) sampling, using a top-hat prior\footnote{Our only constraint here is that the values of $c_1, ~c_2$ and $\kappa$ cannot be such that the resulting $\sigma_{\chi_{\vec{v_{CO}} - \vec{B}}}$ is negative across any projection angle, as this would be unphysical.} and a Gaussian likelihood to estimate the values of $c_1, ~c_2$, and $\kappa$. We find that $c_1 = -8.8 \pm 0.4$, $c_2 = 69 \pm 3$, and $\kappa = 0.37 \pm 0.02$. The resulting fit is shown with the red thick line in Fig.~\ref{SigmafPrj} whereas the thin red lines show 100 random samples from the posterior distributions of the parameters. Therefore, given the values for $c_1, ~c_2$, and $\kappa$ and the measured $\sigma_{\chi_{\vec{v_{CO}} - \vec{B}}}$ in an observed collapsing cloud or core the inclination angle between the LOS and the mean component of the magnetic field can be estimated.

\subsubsection{Determining the projection angle of a new cloud}\label{benchmarking}

We now focus on the inverse problem in order to benchmark this new approach for finding the projection angle. We examine an earlier evolutionary stage of the cloud when the ``central'' $\rm{H_2}$ number density is $2.5\times10^4~\rm{cm^{-3}}$, almost an order of magnitude smaller than the time analyzed in the previous sections and in \greenhyperref{paperI}{Paper I}. At this stage, the cloud has a different magnetic-field morphology, different chemical properties and a different velocity field. We produced new synthetic polarization and spectral-line observations (as described in Sect.~\ref{mockdustmeth} of the present paper and Sect.~2.4 of \greenhyperref{paperI}{Paper I}) under a random projection angle of $\phi = 54.5^\circ$. Additionally, to try to complicate matters somewhat more, we consider a smaller spatial resolution by a factor of two compared to the PPV cubes shown in Fig. 3 of \greenhyperref{paperI}{Paper I} (see also Fig.~\ref{FigVecs}). Noise was also added to the spectra such that the S/N in the central regions of the cloud was equal to 20. We then followed the exact same procedure described in Sect.~\ref{projangleBfield}; namely, we smoothed the data with a Gaussian kernel, with a standard deviation of one pixel, and measured the direction of the ``flow'' between successive velocity slices using the Lucas-Kanade optical flow estimator. While measuring the direction of the ``flow,'' we also required that emission in velocity slices ``n'' and ``n+1'' to be 2.5 times the rms noise. Finally, we compared the orientation of the resulting vectors to the polarization vectors. 

\begin{figure}
\includegraphics[width=1.\columnwidth, clip]{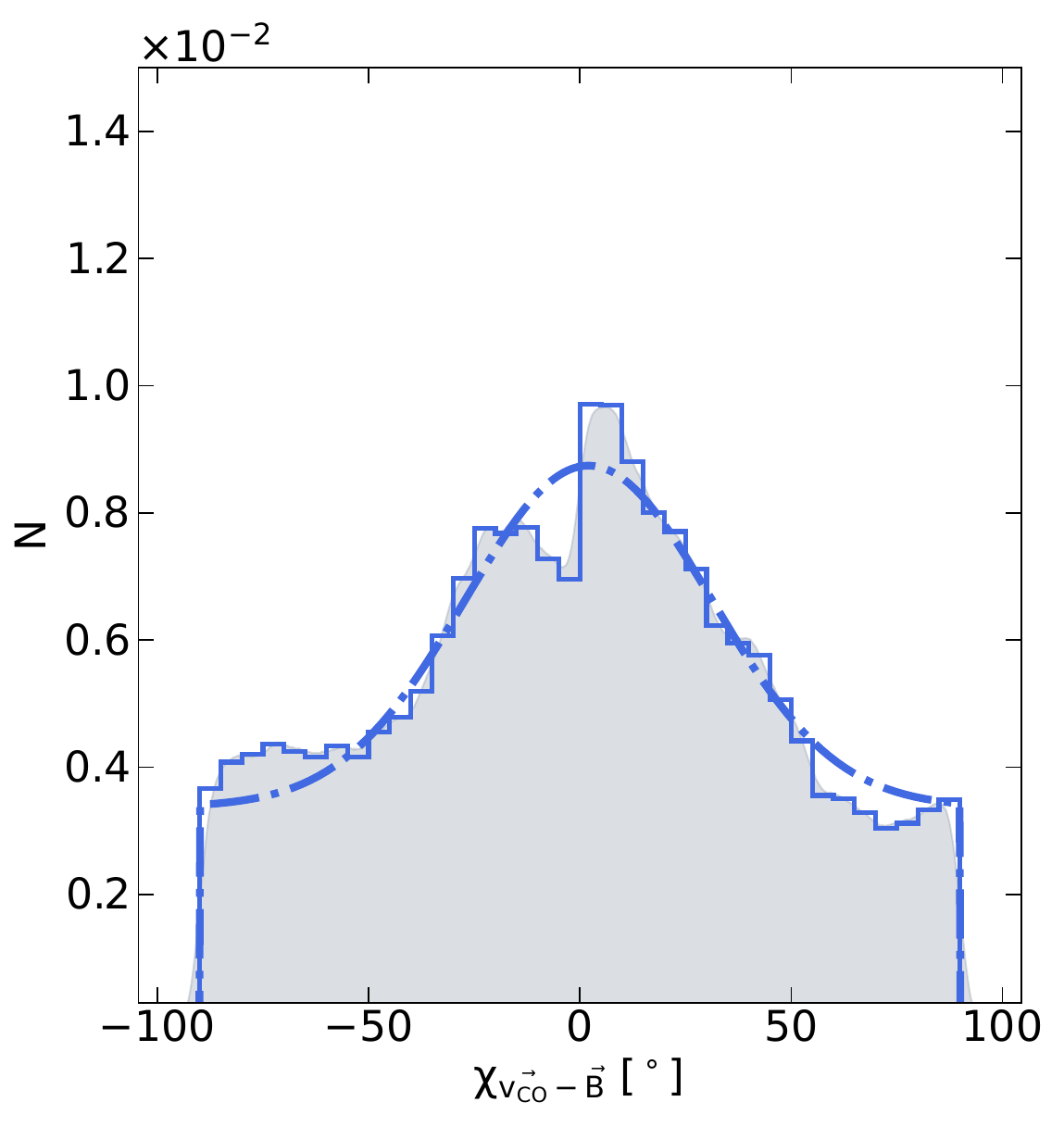}
\caption{Same as the bottom row of Fig.~\ref{AngleDists}, but when the cloud is observed at an earlier evolutionary stage, with a smaller spatial resolution (by a factor of two compared to Fig.~\ref{AngleDists}) and for a different projection angle of $\phi = 54.5^\circ$.
\label{EarlierStageDiffAngle}}
\end{figure}

The resulting distribution of angles between the vectors tracing the direction of the ``flow'' and the pseudo-polarization vectors, together with a fit to the data, is shown in Fig.~\ref{EarlierStageDiffAngle}. By measuring the standard deviation of the fitted Gaussian and using the inverse model (e.g., $\phi = ((\sigma_{\chi_{\vec{v_{CO}} - \vec{B}}} - c_2)/c_1)^{1/\kappa}$), we find that the polar angle is equal to $\phi = 56 \pm 12^\circ$. The uncertainty in the projection angle was calculated by performing a standard Gaussian error propagation as
\begin{equation}
\delta \phi = \sqrt{\left(\frac{\partial \phi}{\partial \sigma_{\chi_{\vec{v_{CO}} - \vec{B}}}} \delta \sigma_{\chi_{\vec{v_{CO}} - \vec{B}}}\right)^2
+ \left(\frac{\partial \phi}{\partial c_1} \delta c_1\right)^2
+ \left(\frac{\partial \phi}{\partial c_2} \delta c_2\right)^2
+ \left(\frac{\partial \phi}{\partial \kappa} \delta \kappa\right)^2}.
\end{equation}
The derived projection angle found using the values for $c_1, c_2$ and $\kappa$ and $\sigma_{\chi_{\vec{v_{CO}} - \vec{B}}}$ is in very good agreement with the true projection angle under which the cloud is ``observed.'' We emphasize here that there was no adjustment or optimization of the parameters $c_1, c_2$, and $\kappa$ in any way or form after they were calculated in Fig.~\ref{SigmafPrj} to achieve this agreement.

\subsubsection{Shortcomings when trying to measure the projection angle}\label{cavs}

Even though the results of the test presented in Sect.~\ref{benchmarking} are promising, there are a number of complications with the analysis proposed here for constraining the projection angle. Firstly, we note that there is one key difference between the thought experiment of a simple disk-like cloud with an hourglass morphology described in Sect.~\ref{projangleBfield} and the results from our simulation. Specifically, in the simplified thought experiment, the angle of the polarization pseudo-vectors should also follow a uniform distribution when such a cloud is observed face-on. However, if we were to compare the vectors tracing the direction of the ``flow'' with the pseudo-vectors we should still get a peaked distribution, as both vectors would point outwards or inwards in a radial fashion from the center of the cloud. 

Given that no such trend is found in our simulations, this points to a situation where the apparent alignment or misalignment between the vectors tracing the direction of the ``flow'' and the polarization is also affected by the turbulent Mach numbers. Therefore, the spread of the fitted Gaussian will also be affected by the turbulent conditions. Additional factors such as rotation, could also further complicate the picture. For instance, if the orientation of the angular momentum is aligned with the magnetic-field orientation and the angular momentum is ``significant enough'' in comparison to gravity, the resulting vectors tracing the direction of the ``flow'' would primarily point perpendicular to the projected magnetic field. However, given the fact that we have not included rotation in the initial conditions of our simulations, we do not currently have a way to quantify how significant this effect might be.

Finally, observational challenges as well as the chemical properties and the evolutionary stage of the cloud might also be factors that affect the direction of the vectors tracing the ``flow'' between successive velocity channels. For instance, if the S/N of the spectra is very low in the outskirts of the cloud or there is no significant emission because the molecule in question is not excited in these regions, the algorithm will only trace the direction of the ``flow'' in a confined region. Therefore, the resulting vectors will not be indicative of the overall ``flow'' in the cloud and their apparent alignment with the polarization pseudo-vectors will also be affected.

Regardless of these complications, given the physical reasons behind the trends observed in Fig.~\ref{AngleDists}, the analysis proposed here has the potential to reveal the projection angle in clouds both on a case-by-case basis and in a statistical fashion. Additional qualitative arguments, such as those described above with respect to how turbulence can affect the apparent alignment between the vectors tracing the direction of the ``flow'' and the polarization pseudo-vectors, could also be used to estimate the turbulent-to-ordered components of the magnetic field \citep{2016JPlPh..82f5301F}.

\section{Summary and conclusions}\label{discuss}

We have produced synthetic dust-polarization observations, under various projection angles with respect to the mean component of the magnetic field, by post-processing an ideal MHD chemo-dynamical simulation. From our synthetic observations, we found that the projection angle has a drastic effect on the observed properties of the cloud. Firstly, even for polar angles that are not far off from exactly ``edge-on'' (e.g., $\phi = 67.5^\circ$), the magnetic field appears significantly more tangled compared to its true 3D structure and the hourglass morphology is not easily identifiable. 

We propose a new approach for analyzing PPV spectral cubes based on the Lucas-Kanade optical flow estimator from the computer-vision community. Based on this algorithm the direction of the ``flow'' between successive velocity channels can be quantified. We show that the direction of the ``flow'' in a collapsing molecular cloud, when viewed ``face-on'' is not correlated with the dust polarization observations. Therefore, the distribution of angles between the vectors that trace the direction of the ``flow'' and polarization vectors resembles a uniform distribution. When the cloud is observed ``edge-on,'' however, there is significant correlation and the resulting distribution of angles resembles the combination of a Gaussian and a uniform distribution. The intermediate projection angles exhibit a gradual transitional behavior between the two extremes. Based on this trend, we have proposed a method to probe the projection angle between the LOS and the magnetic field. First, we simultaneously fit a Gaussian and a uniform distribution to the distribution of angles between the vectors that trace the direction of the ``flow'' and polarization vectors. We then correlated the standard deviation of the Gaussian to the inclination angle. We benchmarked our new approach by performing additional radiative-transfer calculations from the same collapsing cloud at an earlier evolutionary stage. From the calculated standard deviation of the Gaussian, we found a projection angle of $56 \pm 12^\circ$, with the true projection angle under which the cloud was ``observed'' being 54.5$^\circ$. Therefore, the recovered angle is within a 25\% level of accuracy of the true value, based on the largest possible 1-$\sigma$ deviation.

\begin{acknowledgements}

We are grateful to K. Tassis for stimulating discussions. We are also grateful to the referee for their constructive feedback and suggestions which helped enhance the clarity and quality of our manuscript. A. Tritsis acknowledges support by the Ambizione grant no. PZ00P2\_202199 of the Swiss National Science Foundation (SNSF). S. Basu was supported by a Discovery grant from NSERC. C. Federrath acknowledges funding provided by the Australian Research Council (Discovery Projects DP230102280 and DP250101526). The software used in this work was in part developed by the DOE NNSA-ASC OASCR Flash Center at the University of Chicago. This research was enabled in part by support provided by SHARCNET (Shared Hierarchical Academic Research Computing Network) and Compute/Calcul Canada and the Digital Research Alliance of Canada. We further acknowledge high-performance computing resources provided by the Leibniz Rechenzentrum and the Gauss Centre for Supercomputing (grants~pr32lo, pr48pi and GCS Large-scale project~10391), the Australian National Computational Infrastructure (grant~ek9) and the Pawsey Supercomputing Centre (project~pawsey0810) in the framework of the National Computational Merit Allocation Scheme and the ANU Merit Allocation Scheme. We also acknowledge use of the following software: \textsc{Matplotlib} \citep{2007ComputSciEng.9.3}, \textsc{Numpy} \citep{2020Nat.585..357}, \textsc{Scipy} \citep{2020NatMe..17..261V} and the \textsc{yt} analysis toolkit \citep{2011ApJS..192....9T}.

\end{acknowledgements}

%
%

\end{document}